\documentstyle[aps,twocolumn]{revtex}

\def\half{\frac{1}{2}}
\def\opone{\leavevmode\hbox{\small1\kern-3.8pt\normalsize1}}
\newcommand{\be}{\begin{equation}}
\newcommand{\ee}{\end{equation}}
\newcommand{\ba}{\begin{eqnarray}}
\newcommand{\ea}{\end{eqnarray}}
\newcommand{\ban}{\begin{eqnarray*}}
\newcommand{\ean}{\end{eqnarray*}}

\newcommand{\one}{\leavevmode\hbox{\small1\normalsize\kern-.33em1}}

\begin{document}

\title{Sundays in a Quantum Engineer's Life}

\author{Nicolas Gisin \\
Group of Applied Physics, University of Geneva\\
20, rue de l'Ecole-de-M\'edecine, CH-1211 Geneva 4, Switzerland}
\maketitle




\vspace{1cm}

\section{I am a Quantum Engineer, but on Sundays I have principles}
{\it I am a Quantum Engineer, but on Sundays I have principles}, John Bell opened his
"underground colloquium" in March 1983, words which I will never forget! What! John
Bell, the great John Bell, presented himself as an engineer!?! one of those people who make
things work without even understanding how they function?!? whereas I thought of John Bell as
one of the greatest theoretician\footnote{today I would add: one of those people who are
unable to make anything work, but think they know why it doesn't function!}.

In March 1983 the {\it Association Vaudoise des Chercheurs en Physique} organized their
annual one-week course in Montana, an excellent combination of ski and physics, on the foundations of
quantum mechanics\cite{AVCP}. For one of these reasons peculiar to the community of people interested
soleny in these foundations, John Bell was invited, but without any time slot for a presentation. With
some friends we, i.e. the PhD students, managed to convince him to give us an evening lecture,
after dinner, while the professors enjoyed the local wine. At first John declined, arguing
that he had not his transparencies, but when we learned that his wife Mary was to join him during the 
week, we had the perfect counter-argument: Mary would bring the transparencies and we would
organize a room and an overhead projector. The talk took place in the basement, the ceiling was 
too low and the students sat on the ground: a perfect underground atmosphere. When John Bell 
started you could "hear the silence": {\it I am a Quantum Engineer, but on Sundays I have 
principles}.

The next few pages are not supposed to be selfcontained, for the physics the reader is
directed to existing literature. These pages simply contain an illustration of my understanding
of John's words\cite{BellSpeakable}.

\section{Quantum cryptography on Sundays}\label{BellIn}
Today, quantum engineers have many good professional opportunities and even more should be
expected in the near future. Nowadays, for example, they work on quantum cryptography\cite{BB84,GisinRMP01}. 
For this purpose they develop techniques to send "single photon pulses" through
telecom optical fibers from a sender, conventionally called Alice, to a receiver called Bob. 
A convenient solution, according to the engineer, consists in
mimicking the single photons by very weak laser pulses, so that the probability that a pulse
contains more than a photon is neglectable. But on a sunny Sunday the physicist says\cite{BLMS00}: 
"weak pulses are not good, because they may contain two photons". So the engineer works out 
a better single-photon source, based on a photon pair source, using one photon as a trigger for
the single photon pulse, see Fig. 1. 
The physicist, impressed by the engineer's ability, likes the idea so 
much that a special issue of the journal of Pretty Rational Letters is edited\cite{QCInnsbruck,QCLAN,QCGeneva}.
However, on the following Sunday, he realizes that there is no need, in principle, to set the 
source on Alice's side\cite{Ekert91}: the source could as well be at the center! This is much more elegant, 
because of higher symmetry.

But then, in this symmetric configuration, there is no longer any physical object transmitted
from Alice to Bob: where does the correlation then come from? and, what then guaranties the
confidentiality? The engineer doesn't care (it is not even clear whether he understands 
the problem). Actually, it is also unclear to the physicist whether there is a problem or not. 
Until he discovers {\it the inequality}\footnote{How did John Bell mention his inequality?
Very simple, he would just say {\it the inequality}!}. This inequality\cite{Bell64} doesn't explain the
correlation observed by the engineer, nor does it say anything about their fragility, hence
about the security of quantum cryptography. The inequality, or more precisely, its violation,
says that the correlation will never be explained by any theory based only on local variables
(local beables in John's word\cite{beables}), i.e. that any description of the "world out
there" must incorporate some nonlocal influcences.

The quantum engineer enjoys his work. He develops better single photon sources (i.e. sources
coming closer to the ideal case), he works on better detectors, higher bit rates,
implements efficient error correction and privacy amplification algorithms, etc. During
that time, the
physicist writes things like $U_1\otimes U_2\Phi^{(+)}=\opone\otimes U_2U_1^t\Phi^{(+)}$
and concludes that the 1-photon and the 2-photon schemes are logically equivalent\cite{bbm}. He
analyses the optimal eavesdropping attack, assuming Eve (the eavesdropper) is limited only
by the laws of quantum physics. Thus, he finds that Alice and Bob are guaranteed to share a 
higher mutual (Shannon) information than Eve (with either Alice or Bob) if and only if
the error rate between Alice and Bob (something the engineer calls QBER: Quantum Bit Error
Rate) is smaller than $\half(1-1/\sqrt{2})\approx15\%$ \cite{Fuchsetal97,CiracGisin97}. 
This result makes the engineer happy,
his experimental results are well below this 15\% threshold: he is on the safe side. The
physicist, however, has mixed feelings: how can it be that this limit, obtained when comparing
Shannon informations with their logarithms, defines precisely the noise limit above which
the inequality can no longer be violated\footnote{The error rate relates to the 2-photon
interference visibility V as follows: QBER=(1-V)/2 (recall that "outside the visibility" Bob
still has one chance out of two to find the correct result). Hence the mentioned QBER threshold
corresponds to a visibility of $1/\sqrt{2}$, wellknown as the threshold for the Bell-CHSH
inequality\cite{chsh}.}\cite{GisinHuttner97}!?! Is it a coincidence? 
or is it deep\cite{ScaraniGisin01}? (This is still an open question!)

\section{Let's assume that the collapse is real}
Several Sundays pass. The physicist is fascinated by the connection between the bound 
derived from the metaphysical assumption of local hidden variables and the security
of his engineer friend's quantum crypto device. Since a few weeks he got interested
in the infamous wave packet collapse, as a possible interpretation the non local correlation.
Suddenly he thinks: "What if the collapse is real? Could it
be that the collapse triggered by Alice's measurement really prepares the state of the photon
flying to Bob?". The physicist knows, of course, that the collapse and the related measurement
problem are notoriously bad questions, since their prediction are indistinguishable from
those of quantum mechanics without any collapse. But, the metaphysical assumption of local
hidden variables led to interesting physics (and interesting engineering), although the main
result is that they do not exist. May be it is worth trying some metaphysical assumptions
about the collapse and see what kind of experiments should be done to test the assumptions?

This was really a nice Sunday, and the physicist thought about testing the speed of what
Einstein called the spooky action at a distance! If it is really Alice's measurement that prepares
Bob's photon at a distance, says the physicist to the engineer, let's carry out Bob's measurement
at precisely the same time as Alice's, so that the nonlocal preparation has no time to operate.
According to this assumption, the nonlocal correlation should disappear when proper timing is
used. If
the correlation remains, then either there is no collapse, or the speed of the spooky action
is faster than the bound set by the timing accuracy.

The engineer likes the challenge of aligning his system such that both measurements take place
"really at the same time". This is far from obvious, knowing that Alice and Bob are connected
by almost 20 km of optical fibers over a straight line distance of more than 10 km!
But, the engineer has heard of relativity and asks:
"In which reference frame should I align the experiment?".
"Well, hum, I do not really know!, admits the physicists, let's try the most obvious choices:
the reference frame in which the Swiss Alps\footnote{i.e. the reference frame in which all the
massive parts of the experiment are at rest (the lab frame, if you prefer).} 
are at rest! And also
the reference frame of the cosmic background radiation (center of mass of the Universe)!".
The engineer smiles, but since it is fun work  he is willing to try the experiment 
(it gives him a break from the task of improving
these photon counters that are so noisy that one needs to cool them, but if they are too cold
then the dead time has to be increased because the after pulses take more time to resorb).
Okay, says the engineer, but 
"What exactly should be aligned? The beam splitters? The detectors? The computers? The observers?".

The physicist is amazed. Now that he dared to consider the assumption that the collapse is real,
so many questions arise! and each hypothetical answer can, in principle, be tested! How to continue?
The following Sunday, our physicist goes for a walk, with his friend David Bohm\footnote{John
Bell always insisted that Bohm's pilot wave model being experimentally undistinguishable from standard
quantum mechanics should be taught to students at the same level\cite{BellBohm}, but ...
who follows this advice?}.
They spoke about the engineer's question: what should be aligned? Clearly, it should be the
device that triggers the collapse. But what could one reasonably assume as the trigger?
After a few minutes of silence, the physicist states: "it must be the detector! That's where 
the irreversible event happens!". "Possibly, replies Bohm, but I bet it is the beam splitters! 
Because it is there that the particle makes its choice"\footnote{Indeed, in Bohm's pilot wave
model, the irreversible choice is made at the beam splitters, in this model the detectors
merely reveal this choice. This influenced Antoine Suarez and Valerio Scarani when they 
developped their proposal\cite{SuarezScarani}.}.

On Monday, the engineer starts aligning the detectors. Indeed, for him aligning the beam 
splitters sounds even stranger than aligning the detectors.

This experiment has really been performed in 1999 in Geneva\cite{RelCollapse}. The referee reports are interesting,
ranging from fascination to desperation\footnote{Notice that John Bell never published
his papers on the foundation of quantum mechanics in regular physics journals! The reason
being that he wanted to avoid these too often sterile discussions with more or less anonymous
referees.}! No doubt that the experiment was a performance (a better
than 5 ps alignment over almost 20 km of fibers: $\approx10^{-7}$ precision). No doubt that
for many physicists the collapse is taboo (but certainly not for John Bell\footnote{When asked for
advice about a research direction on the foundation of quantum mechanics, John Bell always
replied: "Do you have a permanent position?". If you have a chance to meet Alain Aspect, ask
him about this!}). No doubt also
that much remains to be done, both on the theory side and experimentally (an experiment with
moving beam splitters is under progress in Geneva).

The experimental result established impressive lower bounds on the speed of the spooky action:
$\frac{2}{3}10^7$ and $\frac{3}{2}10^4$ times the speed of light in the "Swiss Alps" 
and the Cosmic Background Radiation
frames, respectively. These numbers are very large, about similar to the ratio between the speed of
sound in air and that of light (for a long time the speed of sound was the fastest measurable
speed, while light was assumed to be instantaneously everywhere).

\section{... and relativity?}\label{bb}
Yet comes another sunny Sunday. The physicist rests in his armchair and thinks:
"All this is quite exciting! But what if we let relativity enter the game even deeper? 
What if the detectors are in relative motion such that each detector in its own reference 
frame analyses its photon before the other?
It would seem then that each photon-detector pair must make their choice before the 
other?!?!?"\cite{SuarezScarani}.

"This renews the tension between quantum physics and relativity", says the physicist
to himself. Indeed the tension is not new. 
Abner Shimony, a good
friend of John Bell, termed this tension as "peaceful coexistence", because the tension
does not lead to any testable conflict\cite{Shimony83}. "However, continues loudly the physicist 
although he is alone, once one assumes that the collapse is a real
phenomena, and once one considers specific models, then the conflict is real and is 
testable\footnote{This comes as a surprise because, until recently, all models were developed
by people whose primary concern was to avoid any testable difference with quantum mechanics:
they were proud that they couldn't be wrong, apparently without realizing that this also
implies that they couldn't be right (i.e. be scientifically relevant\cite{Popper}). 
Very strange indeed!}!".
His idea is that the reference frame in which the collapse propagates is not the Swiss Alps frame, 
nor the cosmic background radiation frame, nor any environmental or universal frame. The intuition is that the frame
is determined by the "trigger device", i.e. by the inertial reference frame of the massive device 
which triggers the collapse. If all trigger devices are at rest in some frame, then it is this
frame which is the relevant one\footnote{If one further assumes that detectors are trigger devices,
then the results recalled in the previous section provide a bound on the speed at which 
entangled objects get separated from a detected one.}. This new assumption opens even clearer
tests: if both measurements happen before the other, then the quantum correlation should
disappear, however large the speed of the spooky action!

Very excited, the physicist starts to evaluate some orders of magnitude. Indeed, the experiment
he just dreamt of requires that one purposely realizes events in moving reference frames such that
the time ordering is altered by relativistic effects. A priori this requires relativistic
speeds. But the formula actually leads to much more optimistic numbers\cite{SuarezScarani}: 
the relative speed $v$ 
should satisfy $v>\frac{c^2\delta t}{\ell}$. With $\delta t\approx 5$ ps and the distance
$\ell\approx 10$ km, one gets $v>50$ m/s. A Ferrari can do it! So, my dear engineer, let's
perform the experiment!!

"But, complains the engineer again, do I really need to activate this bloody moving detector! 
This requires liquid nitrogen and is a real mess!".
The physicist tries to convince the engineer to use "real detectors". But he has to admit 
that he can't explain why: if the assumed collapse happens on the first microns of the 
detector, it should make no difference whether the detector is activated or not\footnote{and
if it takes place later in the detector, then the time jitter (tens or even hundreds of ps)
makes the experiment not yet feasible.}!?! 
Moreover, the effect could be seen anyway: let's arrange the experiment such that one of Bob's
detectors is clearly the first one to interact with the photon. Either this detector detects
the photon and then the second detector will not see the photon. Or the first detector triggers
a collapse corresponding to "no photon" and then the photon gets localized in the path to the 
second detector. Hence, the engineer and physicist compromise: a first experiment 
without active detectors is planned.

A week later the physicist realizes that the unactivated detector was even replaced by a simple
absorber mounted on a rotating wheel. The engineer is happy: now the experiment is feasible! Well, thinks the physicist, this
looks too simple. But what could be wrong?

This experiment was also performed in Geneva, in the spring of 1999\cite{RelCollapse}. The 2-photon interferences
were still visible, independently of the relative velocity between Alice and Bob's reference
frames (actually the magnitude of the velocity was constant at 100 m/s, 
but its orientation varied). What would 
John Bell have thought of this?

\section{Conclusion}
The main lesson I learned from John Bell led me, on the one side to become a quantum engineer,
and, on the other side, not to forget about the principles. In this way one can make a (very)
good living. Simultaneously one can keep ones fascination for the basic questions 
without loosing ground in endless
metaphysical discussions: there is nothing wrong with metaphysical assumptions, but the good ones are
those that can be tested\cite{ExpMeta}. For example, wouldn't it be nicer to dispute the collapse of the 
wave packet as a 
physical phenomenon by designing and performing experiments,
rather than arguing that the collapse is a metaphysical assumption?

\section*{Acknowledgments}
Work supported by the "Fondation Odier de psycho-physique" and the Swiss NSF.
The author is grateful to Cheryl Dotti and Wolfgang Tittel for careful reading
of the manuscript.

\section*{Figure Captions}
\begin{enumerate}
\item{Schematics of the relation between the setups using faint laser pulses (top) 
and those used to test the inequality (bottom). In the upper scheme, each 
pulse has a mean photon number $\mu\approx0.1$. In the second scheme, the source is
replaced by a 2-photon ($2~h\nu$) source, one is used as a trigger, the other is prepared
according to the setting $\alpha$ and sent to Bob. In the third scheme, Alice's photon
is used to prepare (at a distance) Bob's photon. Finally, the last scheme is completely
symmetric between Alice and Bob.}
\item{Schematic of the Geneva long-distance quantum correlation experiments. When the 4
APDs (photon counters) are connected, this establishes a quantum channel for key 
distribution (i.e. quantum cryptography). The same setup allows to test - and violate - the
Bell inequality\cite{Tittel98}. When APD 1 and 2 are set precisely at the same optical distance from the
source, bounds on the speed of the spooky action (nowadays called quantum information!) can be
set, as discussed in section 3. When the APD 2 is replaced by the absorber on the fast rotating
wheel, 2-photon interferences can be observed between APD 3 and 4. This realises the before-before
experiment presented in section \ref{bb}.}
\item{Quantum nonlocality is central to today's physics! Logically it follows directly
from the supperposition principle. This principle leads also to the measurement problem, a
dead end since not testable (at least in the foreseeable future). Besides Quantum Nonlocality
are Relativity and Information Theory, the two other main scientific achievements of the first 
half of the $20^{th}$ century. Relativity is characterized by determinism, information theory
by classical probabilities. From quantum nonlocality and the relativistic no-signaling
condition one can derive the linearity of quantum dynamics, 
hence the Schr\"odinger equation\cite{GisinHPA}.
From quantum nonlocality and information theory one derives quantum cryptography, whose
security is intimately connected to Bell's inequality (section \ref{BellIn}). Finally, the optimal
quantum cloning machine can be defined as the eavesdropping strategy on a quantum
cryptographic channel providing Eve with optimal Shannon information and can be derived from
the no-signaling condition and the existence of distant entangled states\cite{GisinQCM98}. }
\end{enumerate}

\end{document}